\documentstyle[epsf]{mn}

\newif\ifAMStwofonts

\newcommand{\simlt}{\lower.5ex\hbox{$\; \buildrel < \over \sim \;$}}
\newcommand{\simgt}{\lower.5ex\hbox{$\; \buildrel > \over \sim \;$}}
\newcommand{\be}{\begin{equation}}
\newcommand{\ba}{\begin{eqnarray}}
\newcommand{\ee}{\end{equation}}
\newcommand{\ea}{\end{eqnarray}}

\ifoldfss
  \ifCUPmtlplainloaded \else
    \NewTextAlphabet{textbfit} {cmbxti10} {}
    \NewTextAlphabet{textbfss} {cmssbx10} {}
    \NewMathAlphabet{mathbfit} {cmbxti10} {} 
    \NewMathAlphabet{mathbfss} {cmssbx10} {} 
  \fi
  \ifAMStwofonts
    \ifCUPmtlplainloaded \else
      \NewSymbolFont{upmath} {eurm10}
      \NewSymbolFont{AMSa} {msam10}
      \NewMathSymbol{\upi}     {0}{upmath}{19}
      \NewMathSymbol{\umu}     {0}{upmath}{16}
      \NewMathSymbol{\upartial}{0}{upmath}{40}
      \NewMathSymbol{\leqslant}{3}{AMSa}{36}
      \NewMathSymbol{\geqslant}{3}{AMSa}{3E}

      \let\leq=\leqslant 
       
    \fi
  \fi
\fi 

\ifnfssone
  \newmathalphabet{\mathit}
  \addtoversion{normal}{\mathit}{cmr}{m}{it}
  \addtoversion{bold}{\mathit}{cmr}{bx}{it}
  \newmathalphabet{\mathbfit} 
  \addtoversion{normal}{\mathbfit}{cmr}{bx}{it}
  \addtoversion{bold}{\mathbfit}{cmr}{bx}{it}
  \newmathalphabet{\mathbfss} 
  \addtoversion{normal}{\mathbfss}{cmss}{bx}{n}
  \addtoversion{bold}{\mathbfss}{cmss}{bx}{n}
  \ifAMStwofonts
    \ifCUPmtlplainloaded \else
      \UseAMStwoboldmath
      \makeatletter
      \new@mathgroup\upmath@group
      \define@mathgroup\mv@normal\upmath@group{eur}{m}{n}
      \define@mathgroup\mv@bold\upmath@group{eur}{b}{n}
      \edef\UPM{\hexnumber\upmath@group}
      \new@mathgroup\amsa@group
      \define@mathgroup\mv@normal\amsa@group{msa}{m}{n}
      \define@mathgroup\mv@bold\amsa@group{msa}{m}{n}
      \edef\AMSa{\hexnumber\amsa@group}
      \makeatother
      \mathchardef\upi="0\UPM19
      \mathchardef\umu="0\UPM16
      \mathchardef\upartial="0\UPM40
      \mathchardef\leqslant="3\AMSa36
      \mathchardef\geqslant="3\AMSa3E

      \let\leq=\leqslant 

    \fi
  \fi
\fi 

\ifnfsstwo
  \DeclareMathAlphabet{\mathbfit}{OT1}{cmr}{bx}{it}
  \SetMathAlphabet\mathbfit{bold}{OT1}{cmr}{bx}{it}
  \DeclareMathAlphabet{\mathbfss}{OT1}{cmss}{bx}{n}
  \SetMathAlphabet\mathbfss{bold}{OT1}{cmss}{bx}{n}
  \ifAMStwofonts
    \ifCUPmtlplainloaded \else
      \DeclareSymbolFont{UPM}{U}{eur}{m}{n}
      \SetSymbolFont{UPM}{bold}{U}{eur}{b}{n}
      \DeclareSymbolFont{AMSa}{U}{msa}{m}{n}
      \DeclareMathSymbol{\upi}{0}{UPM}{"19}
      \DeclareMathSymbol{\umu}{0}{UPM}{"16}
      \DeclareMathSymbol{\upartial}{0}{UPM}{"40}
      \DeclareMathSymbol{\leqslant}{3}{AMSa}{"36}
      \DeclareMathSymbol{\geqslant}{3}{AMSa}{"3E}

      \let\leq=\leqslant 

    \fi
  \fi
\fi 

\ifCUPmtlplainloaded \else
  \ifAMStwofonts \else 
    \def\upi{\pi}
    \def\umu{\mu}
    \def\upartial{\partial}
  \fi
\fi

\title{Type~Ia supernovae and the formation history of
early-type galaxies}
\author[Ferreras \& Silk]
{Ignacio Ferreras and Joseph Silk\thanks{\tt ferreras,silk@astro.ox.ac.uk}\\
Physics Dept. Denys Wilkinson Building, 
1 Keble Road, Oxford OX1 3RH, United Kingdom}
\date{Draft version \today}
\date{Submitted March 18, 2002; Revised June 11, 2002; Accepted July 17, 2002}
\pagerange{\pageref{firstpage}--\pageref{lastpage}}
\pubyear{2002}

\begin{document}

\maketitle

\label{firstpage}

\begin{abstract}
Using the standard prescription for the rates of supernovae 
type~II and type~Ia, we compare the predictions of a simple
model of star formation in galaxies with the observed 
radial gradients of abundance ratios in a sample of early-type 
galaxies to infer the relative contribution of each type
of supernova. The data suggests a correlation between the
fractional contribution of Type~Ia to the chemical enrichment
of the stellar populations ($1-\xi$) and central velocity 
dispersion of order $1-\xi\sim -0.16\log\sigma_0+0.40$, so 
that the type~Ia contribution in stars ranges from a negligible 
amount in massive ($\sigma_0\sim 300$ km s$^{-1}$) galaxies
up to $10\%$ in low-mass ($\sim 100$ km s$^{-1}$) elliptical 
galaxies. Our model is parametrized by a star formation timescale
($t_{\rm SF}$) which controls the duration of the starburst. 
A correlation with galaxy radius as a power law 
($t_{\rm SF}\propto r^\beta$) translates into
a radial gradient of the abundance ratios. 
The data implies a wide range of formation scenarios 
for a simple model that fixes the luminosity profile, 
ranging from  inside-out ($\beta=2$), to  outside-in formation ($\beta =-1$), 
as is consistent with  numerical simulations of elliptical galaxy 
formation. An alternative scenario that links $t_{\rm SF}$ 
to the dynamical timescale favours  inside-out formation
over a smaller range $0.4<\beta < 0.6$. In both cases, 
massive galaxies are
predicted to have undergone a more extended period of
star formation in the outer regions with respect to 
their low-mass counterparts.
\end{abstract}


\section{Introduction}
Type~Ia supernovae (SNIa) describe stellar explosions whose spectra
show lines of elements or intermediate mass such as silicon, and
of the iron group, but no hydrogen lines. They currently occupy 
a very special place in cosmology since they can be used as
standard candles. Even though their luminosities span an order of 
magnitude, empirical correlations between their absolute magnitude 
and the shape of their light curves can be used in order to determine 
distances (Phillips 1993) in an analogous way to classical 
Cepheids. Furthermore, the absolute luminosity of SNIa is orders of 
magnitude brighter than variable stars, enabling 
us to observe them at high redshift. This is a 
tool currently being used to constrain the cosmological parameters
(e.g. Perlmutter et al. 1999; Riess et al. 1998). However, SNIa
also play a very important role in the process of galactic chemical enrichment 
since they contribute a large amount of iron to the interstellar 
medium (Thielemann, Nomoto \& Yokoi 1986). In fact Type~Ia supernovae 
might even be the main producers of iron in the Universe 
(Ishimaru \& Arimoto 1997; however see Gibson, Loewenstein \& Mushotzky 1997). 
The observed properties of SNIa's suggest a binary model in which at 
least one of the stars is a white dwarf that  reaches the
Chandrasekhar limit ($\sim 1.4M_\odot$) by accretion or by 
merging with another white dwarf. The timescale for the 
explosion is thereby 
limited by the lifetimes of stars which end up as white dwarfs,
i.e. with masses $M\simlt 8M_\odot$.
This implies  that Type~Ia's
can be observed long after star formation subsided. Indeed,
all supernovae observed in early-type galaxies --- which feature
no ongoing star formation --- are Type~Ia, 
whereas late-type galaxies display a mixture of Type~Ia and
Type~II supernovae (Cappellaro et al. 1997)

On the other hand, Type~II supernovae (SNII) show hydrogen lines in 
their spectra and arise from the core collapse of a single,
massive ($M\simgt 8M_\odot$) star at the end of its lifetime,
which occurs between $1-50$~Myrs after the core hydrogen 
burning phase started. Hence, the timescales
of either type of supernovae are remarkably different.
Furthermore, the yields of chemical elements are also in
sharp contrast, since SNIa produce much more iron than
SNIIs, so that stars born during the first phases of
star formation --- when the contribution from SNIa to the
interstellar medium was negligible --- display an 
enhancement of $\alpha$ elements over iron, with respect to the 
younger generations of stars, such as the Sun, which are born 
in an environment polluted by both types of supernovae.

Observations of radial gradients in the colours 
(Franx, Illingworth \& Heckman 1989; 
Peletier et al. 1990; J\o rgensen, Franx \& Kj\ae gaard 1995) 
and line indices (Gonz\'alez 1993; Davies, Sadler \& Peletier 1993;
Peletier et al. 1999) in elliptical galaxies show gradients, 
being mostly redder and more metal rich
at their centres, although some early-type
galaxies display blue cores (Menanteau, Abraham \& Ellis 2001).
Broadband photometry has been a technique repeatedly used to infer
the star formation history of galaxies 
notwithstanding the age-metallicity degeneracy (Worthey 1994),
which prevents us from getting a well-defined picture of galaxy
formation using colours alone. A combined analysis of line indices 
seem to break the degeneracy since their age and
metallicity dependence can be rather different 
(Kuntschner 2000; Trager et al. 2000a, 2000b).
Abundance ratios represent an alternative observable since the
timescales for SNIa and SNII are remarkably different. 
Giant early-type galaxies feature an overabundance of $\alpha$
elements over iron (Peletier 1989; Worthey, Faber \& Gonz\'alez 1992;
Trager et al. 2000a; Kuntschner 2000),  indicative 
of a short duration of the star formation stage, so that mostly
SNII contribute to the metallicity of the stellar component.

This has been used in several models of galaxy formation
in order to constrain the star formation history of galaxies. 
Matteucci (1994) analysed the observed
[Mg/Fe] ratios in ellipticals to constrain the star formation
history to timescales shorter than $0.1$~Gyr. Furthermore, the trend
of [Mg/Fe] with galaxy mass was used to imply either an increasing 
star formation efficiency with galaxy mass, or a top-heavy initial 
mass function, so that more massive stars were formed in 
ellipticals compared to a more quiescent environment such 
as in a disk galaxy. A detailed analysis of abundance ratios in 
metal-poor stars is a valuable tool for  determining the star 
formation history in the solar neighbourhood (Gratton et al. 2000), 
and this can be extended to stellar populations in globular clusters. 
In a recent analysis of abundance ratios in a sample of six
red giant stars in the $\omega$~Cen cluster, Pancino 
et al. (2002) found evidence for the contribution of SNIa
to the composition of younger, more metal rich red giants.

In this paper we explore the radial gradients observed in  
abundance ratios such as [Mg/Fe] in elliptical galaxies as a
function of the star formation timescale. The relative contribution
from SNIa and SNII can be compared with  observations to infer 
the formation process of the stellar components, enabling us
to connect the star formation history and the dynamical history
of early-type galaxies.
In \S2 and \S3 we describe the model used to predict the rates of
either type of supernovae and their contribution to chemical
enrichment. \S4 describes our model predictions and compares
them to observed data. Finally, in \S5 we discuss the implications
of the comparison between our simple model and the observed data.

\section{Rates of type~Ia and type~II supernovae}
The progenitors of type~Ia supernovae are still a matter of debate.
The most common candidates are either the double degenerate
model (Iben \& Tutukov 1984) in which two C-O white dwarfs merge, 
reaching the Chandrasekhar mass and exploding by C deflagration, 
and the single degenerate model (Whelan \& Iben 1973) comprising 
a close binary system made of a nondegenerate star and a 
C-O white dwarf which accretes material from the other component,
reaching the Chandrasekhar mass and triggering C deflagration.
The negative result in the search for close binary systems 
made up of two massive white dwarfs in a sample of 54 white dwarfs
(Bragaglia et al. 1990) hints at the single degenerate model as 
the most plausible one, which is the one we will assume henceforth.
The ESO SNIa progenitor survey, which will use UVES/VLT 
(Koester et al. 2001) on a sample of 1,500 white dwarfs 
will clarify this point.

In order to estimate the rates of type~Ia supernovae, we follow the
prescription of Greggio \& Renzini (1983), recently reviewed by 
Matteucci \& Recchi (2001). The rate can be written as a convolution
of the initial mass function (IMF) over the mass range that can 
generate a type~Ia supernova. The lower mass limit is $M_{Bm}=3M_\odot$
in order to generate a binary with a white dwarf which will reach the 
Chandrasekhar limit. The upper mass limit is $M_{BM}=16M_\odot$
so that neither binary undergoes core collapse. The rate is thus:
\be
R_{Ia}(t)={\cal A}\int_{3M_\odot}^{16M_\odot}dm\phi (m)
\int_{\mu_m}^{\mu_{\rm max}}d\mu^\prime f(\mu^\prime )\psi (t-\tau_{m_2}),
\ee
where $m_t={\rm max}(3M_\odot,m(t))$ and $m(t)$ is the mass of a star
whose lifetime is $t$, whereas $\tau_{m_2}$ is the lifetime of the
nondegenerate companion, with mass $m_2$. $f(\mu )$ is the fraction of binaries
with a mass fraction $\mu\equiv M_2/M_B$, where $M_2$ and $M_B$
are the masses of the secondary star and the binary system,
respectively. The range of integration goes from 
$\mu_m={\rm max}(m_2/m,1-8M_\odot/m)$, to $\mu_{max}=0.5$. 
The analysis of Tutukov \& Yungelson (1980) on a sample of about 1000
spectroscopic binary stars suggests that mass ratios close
to $\mu=1/2$ are preferred, so that the normalized distribution 
function of binaries can be written: 
\be
f(\mu ) = 2^{1+\gamma}(1+\gamma )\mu^\gamma, 
\ee
as suggested by Greggio \& Renzini (1983), and we adopt the value 
of $\gamma =2$. The normalization constant ${\cal A}$ is 
constrained by the ratio between type~Ia and type~II supernovae 
in our Galaxy
that best fits the observed solar abundances. We use the
result of Nomoto, Iwamoto \& Kishimoto (1997), namely
$R_{Ia}/R_{II}=0.12$ to find ${\cal A}=0.05$.

The rate of type~II supernovae is also given as an
integral of the star formation convolved with the IMF,
in this case over the mass range expected for the precursors
of single component supernovae, i.e. $8<M/M_\odot<50$. Here we adopt
$50M_\odot$ as the upper mass cutoff. We use throughout the paper
either a Salpeter (1955) or a Scalo (1986) IMF.
\be
R_{II}(t)=\int_{8M_\odot}^{50M_\odot}dm\phi (m)\psi (t-\tau_m),
\ee
where we have neglected the contribution from binaries to the
IMF over this mass range. 

\begin{figure}
\epsfxsize=3.5in
\begin{center}
\leavevmode
\epsffile{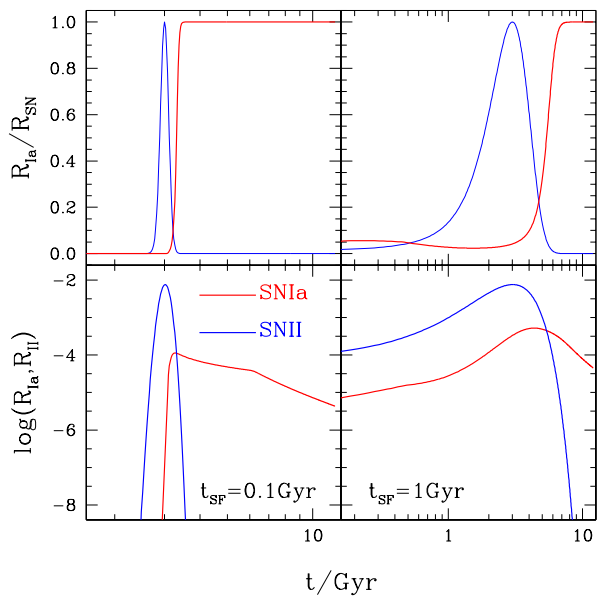}
\bigskip
\caption{Relative contribution ({\sl top}) and rates ({\sl bottom}) 
of type~Ia and type~II supernovae for a generic star formation history 
described by a gaussian function peaked at $3$~Gyr with a star formation
timescale of $t_{\rm SF}0.1$ ({\sl left}) and $1$~Gyr, respectively.}
\end{center}
\end{figure}

Figure~1 shows the rates of either type of supernovae for a generic
star formation history given by a gaussian profile with 
a width of $t_{\rm SF}=0.1$~Gyr ({\sl left panels}) or
$1$~Gyr, respectively. The shape of the curve plotting the 
rate of type~IIs is nearly identical to the star formation rate
given the very short lifetimes of massive stars that undergo 
core collapse. Hence, the rate of this type of supernovae is
negligible a few Myr after the star formation has ceased.
On the other hand, the longer timescales of the nondegenerate
companions in type~Ias imply  that these supernovae start contributing
significantly to the chemical enrichment of the ISM a few hundred Myr after the main burst of star formation. This effect is
more extreme in sharply peaked bursts of star formation although
its contribution to chemical enrichment is only important for the
interstellar medium and not so much for the metallicities of the
stellar populations unless later bursts of star formation 
allow the ejected metals from SNIas to be locked into subsequent
generations of stars.

\section{Chemical enrichment}
The supernova rates allow us to infer the amount of metals ejected
to the interstellar medium. We target magnesium and iron as the
main elements which should be observed to discern the contribution
between SNIa and SNII. Magnesium is one of the $\alpha$ elements,
which are those that can be obtained from the reaction of C 
and O with helium nuclei. These elements are:
Ne, Mg, Si, S. The solar abundance of these elements is 
slightly enhanced with respect to the mean (given by some
average metallicity ${\cal Z}$). Another family of elements --- the 
so-called iron-peak elements, which include Fe and Co --- 
are less abundant with respect to ${\cal Z}$. Notice that 
the amount of iron-peak elements contributes only a small fraction
to the total metallicity (8\% at solar abundance), so that
changing the abundance of Fe by a large amount does not affect
significantly the average metallicity. Hence, it would be more
correct to think of a {\sl depression} in Fe-peak elements in 
solar abundances, rather than an enhancement of $\alpha$ elements
at the centres of elliptical galaxies (Trager et al. 2000b).

The yields from SNIa are taken from the single degenerate model W7 of
Thielemann et al. (1986), which assumes a progenitor made up 
of a $1M_\odot$, C$+$O white dwarf with equal parts of $^{12}$C 
and $^{16}$O, with an accretion rate of 
$4\times 10^{-8} M_\odot$ yr$^{-1}$. 
The yields from SNII are taken from Thielemann, Nomoto \& Hashimoto (1996) 
for a range of progenitor masses. The stellar initial mass function 
(IMF) is used in order to take a weighted-average of the yields with 
respect to mass. The yields used in this paper are shown in Table~1.
Notice that a very significant amount of iron is produced in SNIa 
with respect to the $\alpha$ elements. The mass contained in O-,
Ne- and C-burning shells is too small compared with the mass in 
the Si-burning zone. This implies SNIa ejecta are dominated by
the products of complete and incomplete Si-burning, i.e. a higher
iron yield compared to the ejecta from core-collapse supernovae. 
However, this is still debatable because the yields may depend
rather sensitively on the initial metallicity. For instance, 
assuming an average metallicity of $0.5 {\cal Z}_\odot$ over the
star formation history of the galaxy will lead to a total reduction
of 25\% in the amount of Fe ejected from SNIa (Thielemann et al. 1986).

\begin{table}
{\bf TABLE 1: Supernovae yields (in $M_\odot$)}
\begin{center}
\begin{tabular}{c|cccc}
 Type  &   Mg    &   $\alpha$ &    Fe   & $[\alpha$/Fe$]$\\
\hline\hline
 SNIa            & $0.016$ & $0.438$ & $0.742$ & $-1.33$\\
 SNII(Salpeter)  & $0.122$ & $3.177$ & $0.094$ & $+0.43$\\
 SNII(Scalo)     & $0.102$ & $2.663$ & $0.105$ & $+0.30$\\
\end{tabular}
\end{center}
\end{table}

In this paper we do not consider the contribution of low and 
intermediate-mass stars ($M\leq 8M_\odot$) to chemical enrichment.
We focus on the contribution from either type of supernova
to the star formation history in early-type galaxies through the 
analysis of Mg and Fe. These elements are not produced in 
stars which do not undergo core-collapse, except for very small
traces of Mg in intermediate mass stars (Marigo, Bressan 
\& Chiosi 1996), which can be neglected for our analysis. 
However, one could
argue that the metallicity dependence of the yields from both
supernova types will introduce a dependence on the contribution
to chemical enrichment from lower mass stars. We do not consider
this dependence in our models but we emphasize this point as a
possible factor to improve upon as more developed theoretical models 
of supernova explosions become available.

Another important factor that needs further work is the 
``starting'' abundance ratios one would expect from 
Population~III stars, the very first stars formed. 
They are assumed to have a very different 
IMF, so that their masses are much higher than subsequent 
generations of stars. The study of the yields from this 
population is still under way, but some preliminary modelling
of high-mass helium cores ($64-133M_\odot$) suggests the 
yields present solar abundances for nuclei with even 
nuclear charge (Si, S, Ar) and significant defficiencies 
in odd-charge nuclei (Na, Al, P) (Heger \& Woosley 2002). 
If the pre-enrichment from Population~III stars is important,
one would have to consider three different contributors to the 
abundance of Mg and Fe.

Assuming only two contributors, namely SNIa and SNII, 
we can find the correlation 
between [Mg/Fe] and the relative contribution of each
type of supernova, quantified by the parameter
\be
\xi\equiv \frac{R_{II}}{R_{II}+R_{Ia}},
\ee
where $R_{Ia}$ and $R_{II}$ are the rates of SNIa and SNII,
respectively. Figure~2 shows [Mg/Fe] as a function of $(1-\xi )$ 
for two different IMFs: Salpeter ($\phi (m)\propto m^{-2.35}$; 
Salpeter 1955) and Scalo ($\phi (m)\propto m^{-2.65}$; Scalo 1986).
The data points are from the detailed analysis of early-type galaxies 
in the Fornax cluster by Kuntschner (2000),
and from the sample of Trager et al. (2000a),
evaluated at two radial positions, $r=R_e/8$ and $R_e/2$. The translation
from line indices to metal abundances is done by matching the data
against a grid of simple stellar populations over a range of ages
and metallicities. The data points
are plotted as a function of central velocity dispersion 
($\sigma_0$, {\sl top axis}), 
which has been chosen so that the model prediction
--- plotted against $(1-\xi )$ --- gives a good fit to the data.
This implies a correlation of order:
\be
(1-\xi) \sim -0.16\log\sigma_0 + 0.43.
\ee
The vertical line at $1-\xi =0$ corresponds to the prediction for
a model in which no SNIa contribute to the enrichment of the
stellar populations. Hence, the contribution of SNIa range from 
a negligible amount for massive galaxies ($\sigma_0\sim 300$ km s$^{-1}$)
up to 10\% for low-mass elliptical galaxies ($\sigma_0\sim 100$ km s$^{-1}$).
The timescales associated to the triggering of SNIa imply 
star formation occured in a very short burst in massive ellipticals,
whereas low-mass galaxies have a slightly more extended period of
star formation. This can be accomodated in the framework of hierarchical
clustering if we assume that the latest merging stages of massive ellipticals
do not give rise to star formation but, rather, involve the merging of
the stellar and gas component.
Solar abundances are obtained for a contribution
of SNIa to the total rate ($1-\xi$) of order 10\%, which should not 
come as a surprise since this is the constraint used in order
to compute the proportionaly constant ${\cal A}$ described 
in \S2. 

\begin{figure}
\epsfxsize=3.5in
\begin{center}
\leavevmode
\epsffile{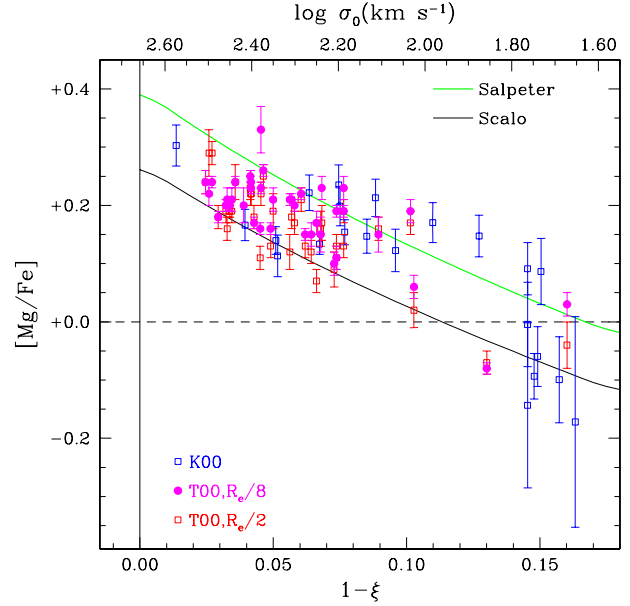}
\bigskip
\caption{Contribution from SNIa to chemical enrichment. 
($1-\xi$) denotes the fractional contribution to the total supernova rate
from SNIa, so that $1-\xi =0$ represents a pure contribution from
core-collapse supernovae. The data points are from Trager et al. 
(2000a; T00), evaluated at $R_e/8$ ({\sl filled squares}) 
and $R_e/2$ ({\sl hollow circles}), and from Kuntschner (2000; K00).
The observed data are plotted against central velocity dispersion 
({\sl top axis}). The model prediction is compatible with the data 
for a linear correlation
between central velocity dispersion and SNIa contribution such as:
$(1-\xi ) = -0.16\log\sigma_0+0.43$.
The steeper slope of the Scalo IMF at the bright end implies a
lower contribution from high mass stars ($M\simgt 20M_\odot$), which
results in a lower enhancement.}
\end{center}
\end{figure}

\section{Abundance gradients}
Observations of colours (Franx et al 1989; Peletier et al. 1990; 
J\o rgensen et al. 1995) and line indices (Gonz\'alez 1993; 
Davies et al. 1993; Peletier et al. 1999) 
in elliptical galaxies show radial gradients, 
being redder and richer in metals at their centres. The observed
spectral indices display a metallicity gradient of
$\Delta [$Fe/H$]/\Delta\log R = -0.2\pm 0.1$ (Davies et al. 1993),
consistent with the observed colour gradients. A metallicity 
gradient can be directly linked to the standard picture of
galaxy formation by the dissipative collapse of a gas cloud.
Carlberg (1984) analyzed such models with an N-body code, 
and predicted abundance gradients around $-0.5$ in 
massive galaxies, flattening towards zero slope in 
lower mass galaxies. Recent merging events in 
massive early-type galaxies will dilute any radial 
signature from an earlier monolithic collapse process, 
thereby flattening the slope (White 1980; Mihos \& 
Hernquist 1994). Hence, abundance gradients allow
us to quantify the degree to which hierarchical 
clustering has contributed to the formation of galaxies.

We assume a simple model with spherical symmetry and a radially-dependent
star formation efficiency. The standard model of star formation
assumes a power-law dependence between the star formation rate 
($\psi$) and the gas density ($\rho_g$; Schmidt 1963), namely:
\be
\psi(t) = C_{\rm eff} \rho_g^n(t)\propto\exp (-t/t_{\rm SF} ),
\ee
where the constant of proportionality ($C_{\rm eff}$) is the 
star formation efficiency, and the exponential function is the
solution for a linear Schmidt law ($n=1$), in which
the star formation timescale $t_{\rm SF}$ can be written
as a function of the efficiency timescale 
($\tau_{\rm eff}\propto 1/C_{\rm eff}$) and the gas 
infall timescale ($\tau_f$) as:
\be
\frac{1}{t_{\rm SF}} = \left( \frac{1}{\tau_{\rm eff}} 
+ \frac{1}{\tau_f} \right).
\ee
Star formation is known to take place in clouds of molecular 
hydrogen. Cloud-cloud collisions contribute to the 
collapse, cooling and subsequent formation of stars. 
In this simple scenario, we would expect the star formation 
efficiency to be proportional to the frequency of cloud-cloud
collisions, which should scale with the number density of
clouds. Hence, the gas density profile traces the star 
formation efficiency, so that at the centre
the star formation timescale should be shorter than in
the outer regions. In this scenario $t_{\rm SF}$ is a 
monotonically increasing function with radius. 

\begin{figure}
\epsfxsize=3.5in
\begin{center}
\leavevmode
\epsffile{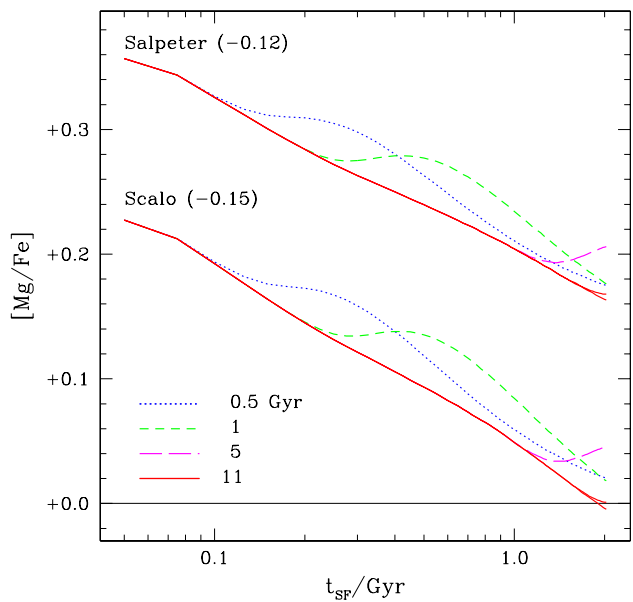}
\bigskip
\caption{[Mg/Fe] profile as a function of the star formation timescale 
($t_{SF}$) which parametrizes the generic star formation history used
in this paper. Several curves are given for different ages of the galaxy:
0.5, 1, 5 and 11~Gyr. The numbers in parenthesis give the slope of 
the profile ($\Delta [$Mg/Fe$]/\Delta\log t_{\rm SF}$) for an age of 11~Gyr.}
\end{center}
\end{figure}

We assume a generic star formation history described by
a gaussian function with a width given by the star formation
timescale ($t_{\rm SF}$), which is left as a free parameter. The
peak occurs at a time $2\times t_{\rm SF}$ and we assume an age of the
Universe of 12~Gyr. Figure~3 shows the predicted relative abundances
of this simple model as a function of $t_{\rm SF}$ for both IMFs
described above and for a set of ages: 0.5, 1, 5 and 11~Gyr. The
number in parenthesis gives the slope of the gradient for both
IMFs. It is beyond the scope of this paper to infer a proper 
functional dependence of the star formation timescale with radius.
Hence, we will just assume a generic power-law dependence:
\be
t_{\rm SF} \propto r^\beta.
\label{eq:beta}
\ee
We find that the predicted abundance radial gradient for old 
stellar ages is $\Delta $[Mg/Fe]$/\Delta r \sim -0.15\beta$. 
In order to compare the model
and the observations, we project the 3D radial coordinate 
($r$) on to a 2D radius ($R$) using a generic 
luminosity profile:
\be
L(<r)\propto r^\gamma \rightarrow 
L(r)\propto r^{\gamma -3}.
\label{eq:lum}
\ee
By projecting the volume distribution on to a surface, we 
can compare the observed slope $a\equiv \Delta$ [Mg/Fe]$/\Delta\log R$
with the model prediction:
\be
a = \gamma - 2 - 0.15\beta.
\ee
Using the deprojected function presented in Hernquist (1990),
which gives a 2D de~Vaucouleurs profile, we can choose $\gamma = 2$. 
This value also matches a NFW profile (Navarro, Frenk \& White 1997) 
for radii shorter than the core radius. In this case, the slope of the
correlation between radius and star formation timescale can
be written:
\be
\beta ({\rm fixed}\ \gamma ) = -\frac{a}{0.15}.
\label{eq:bfg}
\ee
On the other hand, we can fix $\beta$ much in the same way 
as most semi-analytic modellers do (e.g. Baugh et al. 1998;
Kauffmann, White \& Guiderdoni 1993) so that
the star formation efficiency ($1/t_{\rm SF}$) scales with the
dynamical timescale $\sqrt{r^3/GM(<r)}$. If we impose a 
mass-to-light ratio that does not change with radial distance
--- to be expected in galaxies formed in short duration 
bursts --- we can use equation (\ref{eq:lum}) to write $M(<r)\propto 
r^\gamma$, which implies:
\be
\beta = \frac{3-\gamma}{2},
\label{eq:ml}
\ee
and in turn  implies:
\be
\gamma = \frac{a + 2.225}{1.075}.
\ee



Our model assumes that the gradient of the abundance ratios
is caused by a formation scenario with different burst 
lifetimes, which depend on the distance to the centre.
One could argue that a large spread of efficiencies in 
different regions of elliptical galaxies would result in 
a rather large age spread. However, the age-sensitive 
Balmer absoption index $H\beta$ does not show any
correlation with radial distance (Davies et al. 1993).
This implies elliptical galaxies do not have a wide age
spread in their stellar populations. Figure~4 shows that
even though we consider a large range of star formation
timescales, our model is in agreement with the data.
We plot in this figure the mass-weighted age of the stellar component
as a function of galaxy age for three different star formation
timescales. The inset gives the predicted absorption in
$H\beta$ as a function of stellar age for simple stellar
populations with three different metallicities from the
latest population synthesis models of Bruzual \& Charlot
(in preparation). One can see that after $3-4$~Gyr, it
is virtually impossible to determine the stellar ages.
This is the age shown as a dashed horizontal line in 
the figure, which implies that for galaxy ages of
8~Gyr, we can explore a range of star formation
timescales up to $t_{\rm SF}\sim 2$~Gyr and still
reproduce the observed lack of correlation between
radius and Balmer index.

\begin{figure}
\epsfxsize=3.5in
\begin{center}
\leavevmode
\epsffile{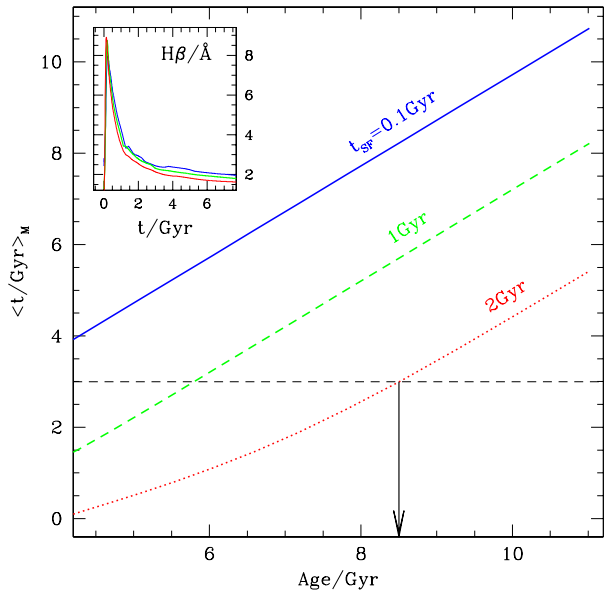}
\bigskip
\caption{Mass-weighted age of the stellar component as a function 
of galaxy age for our generic star formation history. The inset shows
the predicted equivalent width of $H\beta$ for several metallicities
(from the top $Z/Z_\odot = \{ 0.5$,$1$, and $2\}$) using the latest 
population synthesis models of Bruzual \& Charlot (in preparation). 
Mass-averaged ages older than $\langle t\rangle_M\simgt 3$~Gyr (arrow) cannot
be discriminated using Balmer indices.}
\end{center}
\end{figure}

\section{Discussion}

Our model links the abundance ratios to the 
star formation timescale, which cannot be directly
observed. Instead, the data available give radial 
gradients. A simple power law, as described in equation 
(\ref{eq:beta}), can be used in order to throw light on the possible
correlation between the infall rate or the star formation
efficiency and the dynamical properties of the galaxy.
Figure~5 compares the observed radial gradients with
the predicted ones, as a function of [Mg/Fe].
The hollow squares represent the data from 
Trager et al. (2000a), measured at two projected radial 
positions: $R_e/8$ and $R_e/2$, which enables us to determine
a gradient. A clear trend is seen towards positive gradients
in more [Mg/Fe]-enhanced (i.e. more massive) galaxies. 
The other data points come from the compilation of 
Kobayashi \& Arimoto (1999) from which
we selected the work of Gonz\'alez (1993, GON); 
Davies et al. (1993, DSP), and Carollo et al. (1993, CDB)
estimated by the authors of the compilation to have
the most reliable data. There is a very large scatter in this 
compilation, with a wide range of 
slopes, both negative and {\sl positive}. In 
agreement with the data from Trager et al. (2000a), a correlation
can be seen so that galaxies with positive slopes tend to
have the largest super-solar abundances, which correspond to
the most massive galaxies as seen in figure~2. The solid line
gives a least-squares fit to the data from Trager et al. (2000a).
The dashed line also shows a similar trend in the compilation 
of Kobayashi \& Arimoto (1999). Estimating abundance ratios 
from line indices is still a very delicate issue especially
because of the difficulty in finding a suitable set of stars
for calibration. Hence, a theoretical approach is called for,
using the computations of Tripicco \& Bell (1995), who recomputed all
of the Lick/IDS spectral indices from a grid of theoretical spectra
and atmospheres with varying abundance ratios. The response functions
to nonsolar ratios is used to correct the standard population synthesis
models of Worthey (1994), calibrated for solar abundance ratios. 
The uncertainties in such translation frmo abundances to line indices
is discussed in Trager et al. (2000a), and an error bar including these
uncertainties is shown in figure~5.

\begin{figure}
\epsfxsize=3.5in
\begin{center}
\leavevmode
\epsffile{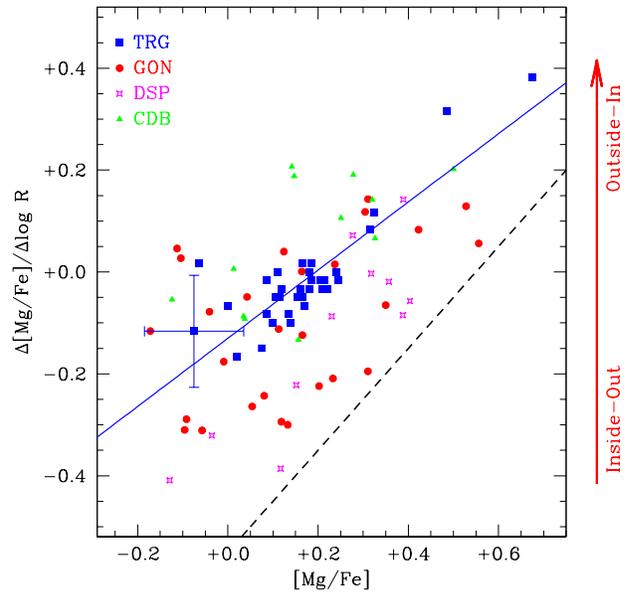}
\bigskip
\caption{
Observed abundance gradients estimated from the
measurements of Trager et al. (2000a; TRG; hollow squares) 
at $R_e/8$ and $R_e/2$. A typical error bar is shown for illustration.
The solid line is a linear squares fit to this data.
A rough estimate of [Mg/Fe] gradients from 
Kobayashi \& Arimoto (1999) is also shown. 
The data points from this compilation correspond to the
sample of Gonz\' alez (1993, GON); Davies et al. (1993, DSP);
and Carollo et al. (1993, CDB). 
The trend clearly seen in the data from Trager et al. (2000a)
as well as hinted at by the void in the lower right corner 
of the data from Kobayashi \& Arimoto (1999), and
illustrated by the dashed line, can be interpreted as a hint towards
a correlation between galaxy mass (which scales with [Mg/Fe]) and
the lower limit of the gradient.}
\end{center}
\end{figure}

A negative gradient implies an inside-out formation process, 
which is suggestive of a monolithic collapse scenario, with
gas falling towards the centres of dark matter halos,
with a gas density profile that decreases outwards and a 
temperature that decreases inwards, thereby generating
a process of star formation that starts at the centre and
spreads outwards. On the other hand, recent TreeSPH numerical 
simulations of galaxy formation (Sommer-Larsen, 
Gotz \& Portinari 2002) find early-type galaxies (both ellipticals
and lenticulars) in which star formation proceeds 
in the opposite way, namely outside-in, mainly triggered 
by merging. Furthermore, recent observations of 
field spheroidals in the Hubble Deep Field (Menanteau et al. 2001) 
show blue cores, result which is currently 
interpreted as secondary bursts of star formation, but which
could be an indication of an outside-in formation process, which
implies a positive gradient. 

In the light of these data, we infer a varying correlation 
between the star formation timescale ($t_{\rm SF}$) and 
galaxy radius. A simple approach, fixing $\gamma$ as in 
equation (\ref{eq:bfg}) gives a slope for this correlation
in the range  $-1.3 < \beta < 2$. A better approximation,
relating $\beta$ and $\gamma$ through the dynamical timescale,
assuming a fixed mass-to-light ratio (\ref{eq:ml}) gives 
$0.4 < \beta < 0.6$ with a 
mass or luminosity profile slope of $1.8 < \gamma < 2.2$.
The higher values of $\gamma$ correspond to positive gradients 
in the abundance ratio.

The scatter is rather large, and it can only imply a very
weak correlation between $\beta$ or $\gamma$ and some dynamical 
parameter such as central velocity dispersion 
or mass. This scatter gives 
another hint of merging as the major mechanism in the assembly
process of early-type systems. In a purely monolithic 
collapse scenario, one would expect a well-defined correlation
between the dynamical timescale and the star formation
timescale, in such a way that most of the early star formation 
would occur at the centre. The void in the lower right corner
of figure~5 (illustrated by the thick dashed line) shows that
we can infer a correlation between [Mg/Fe] (or galaxy mass
as shown in Figure~2) and a lower limit to the gradient.
This can be interpreted as monolithic collapse being possible only
in low mass systems. Higher mass galaxies must be assembled 
through merging. We emphasize here that the inside-out versus
outside-in classification may be an oversimplification. Depending on the
simple model we choose --- i.e. fixed luminosity profile, 
as in (\ref{eq:bfg}), or correlating $\beta$ with the dynamical 
timescale, as in (\ref{eq:ml}), massive galaxies give respectively 
real outside-in star formation ($\beta <0$) or a shallower slope 
for the correlation between star formation timescale and radius.
In either case, the correlation suggested by the dashed
line in figure~5 hints at more star formation 
in the  outskirts of massive galaxies compared to low mass systems 
during the major stage of star formation. 

It is worth mentioning that the slope of the abundance
ratios is a more robust estimator than the absolute abundance.
The latter is strongly dependent on the amount of gas
ejected in outflows, which is an important mechanism
in early-type galaxies as hinted at by the mass-metallicity
relation (Arimoto \& Yoshii 1987; Ferreras \& Silk 2001),
whereas the slope would depend on the scaling of the
outflows with radial distance, which should be similar
to the scaling of the star formation efficiency and 
infall timescale, thereby reducing the effect of outflows
on the slope.

The radial gradients in abundance ratios represent 
an alternative way of studying the universality of the IMF, 
compared to analyses of 
the metallicities of local stars or the intracluster medium 
(Wyse 1997). Figure~3 shows that a non-universal IMF would 
translate into a slope change of the abundance ratio [Mg/Fe]. 
For instance, if we expect 
a top-heavy IMF in environments with a high star formation 
rate (i.e. with a short $t_{\rm SF}$) then the correlation
between [Mg/Fe] and star formation timescale should be steeper.
However, this calls for a more detailed model which is beyond the
scope of this paper.
We have also neglected the effect of early galactic winds 
which could eject $\alpha$-enhanced material out of the galaxy.
A correlation of these winds with the local escape velocity
could explain the observed radial gradient of Mg abundance
in ellipticals (Martinelli, Matteucci \& Colafrancesco 1998).

The model presented here relies on the fact that we understand
the mechanisms that trigger SNIa and we are able to predict their
rates or at least their ratio with respect to SNII. Short bursts
of star formation imply that most of the contamination from SNIa
will go to the ISM and not so much to the stellar component. Hence,
we expect gas in ellipticals to have solar abundance ratios.
There has been a long controversy over this point. The study of
{\sl ASCA} observations of a few clusters hinted at 
a significant SNIa contamination of the gas in rich clusters 
(Ishimaru \& Arimoto 1997). However, Gibson et al. (1997) showed
that a different calibration could invalidate this hypothesis.
Arimoto et al. (1997) even raised the doubt of whether abundance
estimates using the standard iron L-line complex in X-rays 
is giving wrong metallicities. 
The latest analysis of six early-type galaxies in Virgo using 
{\sl ROSAT} and {\sl ASCA} observations (Finoguenov \& Jones 2000) 
seem to agree with a high iron content, i.e. an important contribution
from SNIa to the enrichment of the ISM in elliptical galaxies. 

In this paper, we conclude that a simple treatment of the
abundance ratios allows us to infer the star formation history
and its connection to the dynamical formation history. The 
data seems to rule out a monolithic or secular formation scenario
for the stellar component of massive ellipticals. On the
other hand, the large scatter in the slope of the radial 
dependence of [Mg/Fe] in low-mass galaxies implies both a merging
and a monolithic ``mechanism'' should be invoked for these systems.
A combined analysis of abundance ratios both in the stellar 
populations and in the interstellar medium will enable us to 
explore the connection between these two histories, and to 
quantify the importance of merging events {\sl both} in the 
star formation and dynamical histories of early-type galaxies.

\section*{Acknowledgments}
IF is supported by a grant from the European Community under 
contract HPMF-CT-1999-00109. It is a pleasure to thank Rosie Wyse
and Scott Trager for very useful comments, Jesper Sommer-Larsen 
for his latest simulations of galaxy formation, and 
Harald Kuntschner for making available his Fornax cluster data.
The anonymous referee is acknowledged for useful comments and 
suggestions.

\end{document}